\def\be{\begin{equation}}
\def\ee{\end{equation}}
\def\ba{\begin{eqnarray}}
\def\ea{\end{eqnarray}}

\def\negenspace{\kern-1.1em}

\def \quabla{\sqcup \kern-9pt \sqcap}

\documentclass{ws-procs975x65}

\begin{document}

%to switch ON running title
\markboth{E.W. Mielke, F.V. Kusmartsev, and F.E. Schunck}{Inflation,
bifurcations of nonlinear curvature Lagrangians and dark energy}

%\wstoc{Inflation, bifurcations of nonlinear curvature Lagrangians and dark
%energy}{E.W. Mielke, F.V. Kusmartsev, and F.E. Schunck}

\title{Inflation, bifurcations of nonlinear curvature Lagrangians and dark energy}

\author{Eckehard W. Mielke}

\address{Departamento de F\'{\i}sica,\\
Universidad Aut\'onoma Metropolitana--Iztapalapa,\\
Apartado Postal 55-534, C.P. 09340, M\'exico, D.F., MEXICO\\
\email{ekke@xanum.uam.mx}}
\author{Fjodor V. Kusmartsev}
\address{Department of Physics, Loughborough University, Loughborough,\\
Leicestershire LE11 3TU, United Kingdom\\
\email{f.kusmartsev@lboro.ac.uk}}
\author{Franz E. Schunck}
\address{Institut f\"ur Theoretische Physik, Universit\"at zu K\"oln, 50923
K\"oln, Germany\\
\email{fs@thp.uni-koeln.de}}

% WARNING. in standard latex cls file formatting, at this point
% \maketitle would typeset the above titlepage information
% but WS has chosen to be nonstandard and have each line typeset
% as it is digested.
% no abstract is necessary.
% \bodymatter below resets the footnote counter and symbols after
% possible use in the title matter.

%\begin{abstract}
A possible equivalence of scalar dark matter, the inflaton, and {\em modified
gravity} is analyzed. After a conformal mapping, the dependence of
the effective Lagrangian on the  curvature is not only {\em
singular} but also bifurcates into several {\em almost Einsteinian
spaces}, distinguished only by a different {\em effective} gravitational strength
and cosmological constant. A swallow tail {\em catastrophe} in the
bifurcation set indicates the possibility for the coexistence of
different Einsteinian domains in our Universe. This `triple
unification' may shed new light on the nature and large scale
distribution not only of dark matter but also on `dark energy',
regarded as an effective cosmological constant, and inflation.
%\end{abstract}

%\bodymatter

%****************************************
\section{Introduction}
The origin of the principal constituents of the Universe in the form
of {\em dark matter} (DM) and {\em dark energy} (DE) (or rather dark
tension due to its negative pressure), remains a major puzzle of modern
cosmology and particle
physics. Ad hoc proposals like modifying Newton's law of gravity as
in Milgrom's MOND (Modified Newtonian Dynamics) are difficult to
reconcile with relativity, or need a phase coupling to a {\em
complex scalar} \cite{Be88}. The particle physics' view is to leave
gravity untouched but postulate WIMPs (Weakly Interacting Massive
Particles) like axions, dilatons or neutralinos  as dark matter
candidates \cite{ZH04}.

The dominant non-visible ``dark" fraction of the total energy of the
Universe is known to exist from its gravitational effects. Since the
{\em dark matter}  part is distributed over rather large distances,
its interaction, including a possible self-interaction \cite{SS00},
must be weak. The main candidates for such weakly interacting
particles  are the (universal) {\em axion} or the lightest
supersymmetric particle, as the neutralino  in most models,
cf.~Ref.~\cite{ZH04}.

Recently, an observed excess of diffuse gamma rays has been
attributed \cite{deB04} to the annihilation of DM in our Galaxy. The
flux of the presumed neutralino annihilation allows a reconstruction
of the distribution of  DM in our Galaxy. Most probable is a
pseudo-isothermal profile but with a substructure of two
doughnut-shape {\em rings} in the galactic plane. It is believed
that these transient substructures have their origin in the hierarchical
clustering of DM into galaxies. However, there are reservations
concerning the internal consistency in the interpretation of the
observations: Given the same normalization for the cross section of
the DM particle, it appears unlikely \cite{An05} that DM
annihilation is the main source of the extragalactic gamma-ray
background (EGB). Moreover, then the large accompanied antiproton
flux needs to meet several constraints \cite{BE06}.

On the other hand, heterotic string theory provides a very light
universal axion which may avoid \cite{GK06} the strong $CP$ problem
in quantum chromodynamics (QCD) \cite{Pe99}. Given the existence of
such almost massless (pseudo-)scalars, it has been speculated that a
coherent {\em non-topological soliton} (NTS) type solution of a
nonlinear Klein-Gordon equation can account for the observed halo
structure, simulating a {\em Bose-Einstein condensate} of
astronomical size \cite{HBG00,JB01,MFS06}; cf.~Ref.~\cite{S98}.  In particular,  a
$\Phi^6$ toy model \cite{MS02} yields {\em exact} Emden type
solutions in flat spacetime, including a flattening \cite{MP04} of
halos  with ellipticity $e<1$ as observed via microlensing
\cite{HY04,SFM06}.

Both views are to some extend physically equivalent, i.e., {\em
scalar} dark matter minimally coupled to Einstein's general
relativity (GR) is equivalent \cite{MS94,bmmov,SO97,SKM05,OUS05} to a {\em modified
gravity} in the relativistic framework of higher-order curvature
Lagrangians. Such effective Lagrangians may also arise from the
low-energy limit of (super-) strings. The   model proposed by Carroll
et al.~\cite{CD03}  also uses such an equivalence  of a nonlinear
higher-order curvature Lagrangian, in order to explain the present
cosmic acceleration, but takes resort to an $1/R$ type curvature
Lagrangian, which is unbounded for weak gravitational fields.

Quite generally, the dependence of the effective Lagrangian $L$ on
the scalar curvature $R$ is  {\em singular}  and cannot always be
represented analytically in the $(R,L)$ plane. More stunning is our
recent finding \cite{SKM05} that, although nonlinear, a $L=L(R)$
type Lagrangian bifurcates in several branches of {\em almost
linear} Lagrangians
\be
 L\simeq \frac{1}{2\kappa_{\rm eff}}\left(R -2\Lambda_{\rm eff}\right)
\label{Eineff}
\ee
distinguished only by the {\em effective} gravitational constant
$\kappa_{\rm eff}$ and a cosmological constant $\Lambda_{\rm eff}$.

This indicates a unifying picture of dark matter, `dark energy'\cite{S98} with
an equation of state parameter $w_{\rm DE}=p/\rho\simeq -1$, and modified
gravity which may account, on different scales, for inflation
\cite{MS95,MP04}, dark matter halos \cite{MS02,MP02} of galaxies or
even dark matter condensations, the so-called boson stars
\cite{kusm,kusm2,MS00,SM03} as candidates of MACHOs (Massive Compact Halo
Objects). The `landscape' view \cite{LU06} of (super-)strings
suggests also such a triple unification based on coherent, possibly
oscillating scalar field with a mass larger than the Hubble scale at the present epoch,
i.e. $m> H_* = 10^{-23}$ eV.

%**************************
\section{Gravitationally coupled scalar fields
\label{trafo}}
Conventionally, dark matter and inflation can be modeled by a
real (or complex) scalar field $\phi$ with self-interaction 
$U(|\phi|^2)$ {\em minimally} coupled to gravity with  the Lagrangian
density
\be 
{\mathcal L}_{\rm DM} = \frac {\sqrt{\mid g\mid
}}{2\kappa } \left \{ R
 + \kappa
   \Bigl [ g^{\mu \nu} (\partial_\mu \phi^*) (\partial_\nu \phi )
             - 2U(|\phi |^2) \right ] \Bigr \} \, ,
 \label{lagrBS}
\ee
where
$\kappa = 8\pi G$ is the gravitational constant in natural units, $g$
the determinant of the
metric $g_{\mu \nu }$, and $R$ the scalar curvature  of
Riemannian spacetime with Tolman's sign conventions \cite{tolman}.
A constant potential $U_0= \Lambda / \kappa $ would simulate the
cosmological constant $\Lambda $.

In a wide range of inflationary models, the underlying dynamics is
simply that of a single scalar field --- the {\em inflaton} ---
rolling in some underlying potential. This scenario invented by
Linde \cite{linde} is generically referred to as {\em chaotic
inflation} due  to its choice of initial conditions. Many
superficially more complicated models can also be rewritten in this
framework.

%******************************************************
\section{General metric of a  flat inflationary
universe}

A spatially {\em flat} ($k=0$) Friedman-Robertson-Walker (FRW) Universe
with metric
\be
ds^2 = dt^2 - a^2(t) \left [ dr^2 + r^2 \left (
       d\theta^2 + \sin^2 \theta d \varphi^2 \right ) \right]
\; \ee
is nowadays favored by observations \cite{Sp03}. Its temporal
evolution of the generic
model (\ref{lagrBS}) is
determined by the two autonomous first order equations
\ba
\dot H & = &- 3H^2(1+w_{\rm DE})= \kappa U(\phi ) - 3H^2 =: V(H,\phi ) \; , \label{doth} \\
\dot \phi & = & \pm \sqrt {\frac {2}{\kappa }}
  \sqrt{3H^2 - \kappa U(\phi )}
 = \pm \sqrt {-\frac {2}{\kappa } V(H,\phi )}
 \; ,\label{dotphi2}
\ea
where $H=\dot  a(t)/a(t)$ is the Hubble expansion rate.

The function $V(H,\phi )$ will turn to be the ``height function" in
Morse theory \cite{miln}. Observe that $V\le 0$ in order to avoid
scalar ghosts. For the FRW metric, the Lagrangian density
(\ref{lagrBS}) reduces to
\be {\mathcal L} = - \frac {3}{\kappa }
\dot a^2 a + \left [
 \frac {1}{2} \dot \phi^2 - U(\phi ) \right ] a^3 \; .
\ee
Since the shift function is normalized to one for the FRW
metric, the canonical momenta are given by $P=\partial {\mathcal
L}/\partial \dot a=-6Ha^2/\kappa $ and $\pi = \partial {\mathcal
L}/\partial \dot \phi = a^3 \dot \phi $. The Hamiltonian or ``energy
function'' is given by
\be E = \Biggl [\frac{1}{2 } \dot \phi^2 +
\frac{1}{\kappa }
    V(H,\phi ) \Biggr ] \, a^3  \cong 0
\ee
and, due to (\ref{dotphi2}), vanishes for all solutions, which is a familiar constraint
in GR.

Using the Hubble expansion
rate $H:=\dot  a(t)/a(t)$ as the {\em new} ``inverse time"
{\em coordinate}
\be
 t  =  t(H) = \int
   \frac {dH}{\kappa \widetilde U - 3 H^2}  \label{tH} \, ,
\ee
we were able to find the general solution \cite{SM94}:
\ba
a & = & a(H) = a_0 \exp \left (
   \int \frac {H dH}{\kappa \widetilde U - 3 H^2} \right )  \label{aH}
\; . \\
ds^2 & = &
\frac {dH^2}{\left (\kappa \widetilde U - 3 H^2 \right )^2}
 - a_0{}^2 \exp \left ( 2
   \int \frac {H dH}{\kappa \widetilde U - 3 H^2} \right ) \times  \nonumber \\
 & &   \left [ d r^2 + r^2 \left (
    d \theta^2 + \sin^2 \theta d \varphi^2 \right ) \right ] \; , \\
\phi & = & \phi (H) = \mp \sqrt {\frac{2}{\kappa }}
 \int \frac {dH}{\sqrt{3H^2-\kappa \widetilde U}} \; ,
\label{phiH}
\ea
where $\widetilde U = \widetilde U (H) := U(\phi(t(H)))$ is the
{\em reparametrized} inflationary potential.

The singular case $\widetilde U = 3H^2/\kappa $ with $w_{\rm DE}=-1$ leads in
(\ref{doth}) to the {\em de Sitter inflation} with  exponential
expansion $a(t) = a_0 \exp (\sqrt{\Lambda /3 \, } t)$, which for a
later time becomes physically unrealistic, since the inflationary
expansion eventually needs to merge into the usual Friedman cosmos.
Therefore, in explicit models we use the ansatz
\be 
\widetilde U(H)
= \frac {3}{\kappa } H^2 + \frac {g(H)}{\kappa} \label{uh}\, ,
\ee
for the potential, where $g(H)= V\left(H,\phi(t(h)) \right) $ is a
nonzero function which should provide the {\em graceful exit} from
the inflationary phase to the Friedman cosmos. In order to have a
positive acceleration during the inflationary phase, but on the
other hand  a real scalar field in (\ref{phiH}), its allowed range
is $-H^{2}<g<0$. This $H$--formalism facilitates considerably the
reconstruction of the inflaton potential \cite{SM94,MS95,MP02}. In
second order this is based on an Abel equation \cite{GM97} for the
primordial density $\epsilon =-g/H^2$.

%********************************
\subsection{Classification by catastrophe theory}

A general classification of
all allowed inflationary potentials and scenarios has already been
achieved by Kusmartsev et al.~\cite{Kus} via the application of {\em
catastrophe theory} to the Hamilton--Jacobi type equations
(\ref{doth}) and (\ref{dotphi2}) regarded as an {\em autonomous
nonlinear system}.

In  phase space, the equilibrium states  are given by the constraint
$\{ \dot H,\dot \phi \}=0$. The critical or equilibrium points,
respectively, of this system are determined by $V(H_c,\phi_c)=0$.
This constraint is {\em globally} fulfilled by $\kappa U(\phi ) =
3H_{\Lambda }^2$, where the Hubble expansion rate is constant,
i.e.~$H_{\Lambda }=:\sqrt{\Lambda /3}$ corresponding to $w_{\rm DE}=-1$. 
For $\dot \phi=0$ and
$\Lambda \neq 0$, we recover the de Sitter inflation.

In general, the Jacobi matrix
\be
J = \left (
\begin{array}{cc}
   -6 H & \qquad \kappa U' \\
   \pm 6 H (- 2 \kappa V)^{-1/2} & \qquad \mp \kappa U'(- 2 \kappa V)^{-1/2}
\end{array} \right )
\; , 
\ee 
of the system (\ref{doth}) and (\ref{dotphi2}) depends
crucially on $U'=dU/d\phi $. Since its determinant vanishes, i.e. $\det J = 0$, the
system is always degenerate. For the analysis of stability, it
suffices therefore to consider
only (\ref{doth}) and  to reconstruct $\phi$ later.
Since $H$ and $\phi$ are independent variables, we can introduce the
non--Morse superpotential $W(H,\phi)$,  defined via $V:=-\partial
W/\partial H$, and analyze the system with the aid of {\em
catastrophe theory}. From (\ref{doth}) we obtain
\be 
W(H,\phi) = H^3
- \kappa U(\phi ) H + C(\phi) \; , \label{whitney}
\ee
where $C$ is
an arbitrary function of $\phi $. The function $W$ is already in
canonical form in $H$--space, and belongs to a Whitney surface\cite{Ar02,Ku89} or to
the Arnold singularity class $A_2$. The
corresponding Whitney surface has only one control parameter, the
potential $U$. Thus, an evolution of critical points is determined
via the values of the potential $U$. Let us analyze the types of
critical points at different fixed values of the control parameter
$U$.

If $U_c :=U(\phi_c)<0$, the equation has no stable critical
point due to the shape of the Whitney surface. However if
$U_c>0$, there are two critical points: a stable one
at $H_c=\sqrt{U_c/3}$ and an unstable one at
$H_c=-\sqrt{U_c/3}$.

The value $U_c=0$ is the {\em bifurcation point}. Provided this is
also an extremal of $V$, we necessarily have $\partial V/\partial H
\mid_c=-6H_c=0$, $\partial V/\partial \phi\mid_c =\kappa U'_c=0$,
and $\dot \phi_c=0$. Thus, also the critical points of the
Klein--Gordon equation for the scalar field are involved. Hence, the
Hubble parameter has to vanish and $\phi_c$ is a double zero of the
potential $U$. The Hessian of (\ref{doth}) takes the form
\be 
{\rm Hess} (V) = \left ( \begin{array}{cc}
 -6 & 0           \\
  0 & \kappa U''
\end{array} \right )
 \; .
\ee
The sub--determinants of the Hessian are $\Delta_{0}=\partial^2
V/\partial H^2=-6<0$ and $\Delta_1= \det {\rm Hess} (V)= (\partial^2
V/\partial H^2) \; (\partial^2 V/\partial \phi^2) - (\partial^2
V/(\partial H \partial \phi ))^2 = -6 \kappa U''$. For a maximum of
the potential $U$, i.e.~$U'=0$ and $U''<0$, the function $V$
possesses a maximum; a minimum of the potential $U$, however,
corresponds to  a saddle point of $V$.
%*************************************************************

\subsection{Reheating}

Since $U(\phi_c)$ can be associated with the {\em latent heat} of the
Universe in this epoch, we could demonstrate \cite{Kus}:

{\em The critical points of the non--Morse potential
$W(H,\phi )$ determine the evolution in the inflationary phase.
Along the minima and maxima $H_c=\pm \sqrt{U(\phi_c)/3}$,
the inflaton moves from the slow--roll
to the hot regime. The saddle points of $W$, i.e.~more precisely, the minima of
$V$, determine the onset of reheating}.

\subsection{Scale-invariant spectrum as a limiting point}
It is now possible to relate the ``slow--roll'' condition, for the
velocity of the inflationary phase, to the critical points resulting
from catastrophe theory. For inflation (with $\ddot a>0$) the
 two ``slow--roll parameters'' are, in first order
approximation,
\be
\epsilon= -g/H^2\quad {\rm and} \quad \eta=
-dg/d(H^2)\;,
\ee
where $g=V(H, \phi)$ is the ``graceful exit function". In this
reduced dynamics, they are effectively determined by the first and
second derivatives of the reduced non--Morse function $W(H)$, i.e.,
more precisely by $\epsilon= (1/H^2)(\partial W/\partial H)$ and $\eta =(1/2H) \partial^2
W/(\partial H)^2$. They will also determine the density fluctuations of the
early Universe.

The deviation $\Delta:= (n-1)/2$ from the scale-invariant
Harrison-Zeldovich spectrum with $n=1$ leads in the first order
slow-roll approximation to the differential equation
\be
H^2\frac{d\epsilon}{dH^2} -\epsilon=\Delta
\ee
with 
\be
\epsilon=
-\Delta +AH^2\,, \quad V= \Delta H^2 -AH^4
\ee 
as solutions for the
density and the graceful exit function, respectively.

Then the non-Morse potential turns out to be \be W(H, \phi)=-\int V
dH =-\frac{\Delta}{3} H^3 +\frac{A}{5}H^5 +C(\phi) \ee which
together with (\ref{whitney}) provides us with the reparametrized
potential 
\be 
\kappa\widetilde U(\phi)
=\left(1+\frac{\Delta}{3}\right) H^2 -\frac{A}{5}H^4\,, 
\ee which
for $\Delta\simeq 0$ is positive in the leading order.

In the  next order slow-roll approximation arises a nonlinear Abel
equation \cite{GM97} with a continuous spectrum for $n<1$ and a
discrete `blue` one for $n>1$, see Ref. \cite{SM00} for more
details. This division of the spectrum persists in the next to
second order slow-roll approximation \cite{MP02}. More important is
our finding that in both higher order approximations the Harrison-Zeldovich
spectrum with $n=1$ stays a {\em limiting point}, which
agrees quite well with recent constraints from the observations of
WMAP \cite{KK06}.

%************************************************************************
\section{Higher--order curvature Lagrangians
via field redefinition}

Nonlinear modifications of the Einstein--Hilbert action are of
interest, among others, for the following reasons: First, some
quadratic models can be renormalized when quantized, cf. Ref.
\cite{MR05}. Second, specific nonlinear Lagrangians have the
property that the field equations for the metric remain second order
as in GR; these are the so--called Lovelock actions  which arise
from dimensional reduction of the Euler characteristics, cf. Ref.
\cite{Mu90}. In Yang's theory of gravity, this topological invariant
induces an effective cosmological constant for instanton solutions
residing in Einstein spaces\cite{Mi81,MR05}. At times\cite{Va02},
this renormalizable model is referred to as {\em Yang-Mielke theory}
of gravity.

As an instructive example of higher-derivative theories of
gravity\cite{SS90}, we consider the Lagrangian density 
\be 
{\mathcal
L}=L(R)\sqrt{\mid g\mid} = \left (\frac{1}{2\kappa} R + \frac{1}{2
\beta^2} R^2 \right) \sqrt{\mid g\mid}, 
\ee 
where  $\beta$ is a
dimensionless coupling constant. Through the conformal change
\cite{Mi77} 
\be 
g_{\alpha \beta} \quad \rightarrow \quad \widetilde
g_{\alpha \beta} = \Omega g_{\alpha\beta}\quad {\rm  with} \quad
\Omega = 2\kappa \frac{dL}{dR}= \left ( 1 + \frac {2\kappa}{\beta^2}
R \right ) \label{confm}\, , 
\ee 
of the metric, this Lagrangian can
be mapped to the usual Hilbert--Einstein Lagrangian with a
particular self--interacting scalar field. Indeed, Starobinsky
\cite{St80} considered earlier such models for inflation.

The scalar field, the inflaton, will arise via 
\be 
\phi =
\sqrt{\frac{3}{2\kappa}}\,  ~\ln \Omega \label{scal} 
\ee 
from the
nonlinear parts of a higher--order Lagrangian $L=L(R)$ in the scalar
curvature $R$. Recently, such modified gravity models are considered
as  alternatives\cite{NO06,CS06} for dark matter or even dark
energy.

\begin{figure}[ht]
\centering \leavevmode\epsfysize=9cm \epsfbox{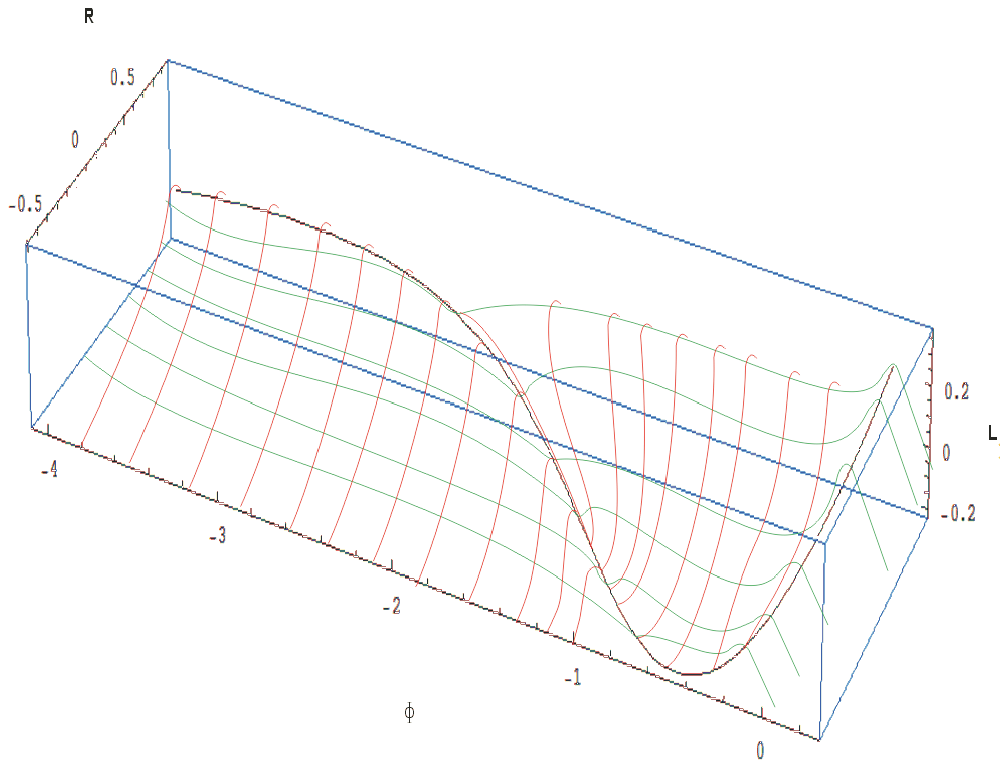}\\ \vskip
0.5cm \caption[fig1]{\label{fig1} The qualitative form of a Whitney
surface with local extrema in the case of a nonlinear curvature
Lagrangian $L(R)$ depending on $\phi$ as control parameter.}
\end{figure}

%********************************************************
\subsection{Reparametrized Lagrangian}
Instead of studying the resulting complicated nonlinear field equations of
higher--order curvature Lagrangians, we follow the equivalence proof of
Ref. \cite{MS94} for the conformal frame (\ref{confm}). Then, our inflaton Lagrangian
density (\ref{lagrBS}) acquires the form
\be
 {\mathcal L} = \frac{1}{2\kappa} \sqrt{\mid g\mid}
\left[ R\Omega -2 \kappa \Omega^{2} {\hat U} (\Omega) \right]\, ,
\ee
where ${\hat U}(\Omega):=
U(\phi(\Omega))=U\left(\sqrt{3/2\kappa} \ln \Omega \right)$ is the
reparametrized potential. Thus in our approach, the inflaton or dark
matter scalar will not be regarded as an independent field,  but is
induced via  (\ref{scal}) by the non--Einsteinian pieces of the
general Lagrangian $L = L(R)$. Solving for the  potential via the
method of Helmholz, we obtain 
\be 
{\hat U} (\Omega) = {\mathcal
H}(R)/\Omega^{2} = \left[ \frac{R\Omega}{2\kappa} -
L\right]/\Omega^{2} \label{6.8}\, . 
\ee

If we identify the conformal factor with the field momentum via
$\Omega= 2\kappa d{L}/dR$, the bracket in (\ref{6.8}) can be
regarded \cite{Mi91} as a {\em Legendre transformation} $L \to
{\mathcal H}(R)=Rd{ L}/dR - {L}$ from the original Lagrangian
(\ref{lagrBS}) to the general nonlinear curvature scalar Lagrangian
$ L = L (R)$. Then,  the {\em parametric reconstruction} 
\be
R=2\kappa \exp \left(\sqrt{2\kappa/3}\phi\right) \left[2
U(\phi)+\sqrt{3/2\kappa}\, \frac{dU}{d\phi}\right] \label{9.4}\, ,
\ee and \be L=\exp \left(2\sqrt{2\kappa/3}\phi\right)
\left[U(\phi)+\sqrt{3/2\kappa}\, \frac{dU}{d\phi} \right ]
\label{9.5} 
\ee
of the higher--order {\em effective}
Lagrangian $L(R)$ from the self--interacting inflaton potential $U(\phi)$
arises \cite{bmmov}.
Here the scalar field
plays merely the role of a control parameter. The form of the Whitney surface
with its local valleys and mountains is qualitatively shown
in  Fig.~\ref{fig1}.

%**************************
\section{Bifurcations with effective
`dark energy' \label{Bifu}}
In order to model self-interacting dark matter, let us consider  the
potential
\be
U = m^2 |\phi |^2\left(1 -\chi |\phi |^4\right)\, ,
\label{Urot}
\ee
where $m$ is the  mass of an ultra-light scalar and
$\chi$ the coupling constant of the nonlinear self-interaction. It
provides us with a solvable non-topological soliton type  model of
dark matter halos \cite{MS02,MSP02} even with toroidal substructures\cite{MV07}
and a reasonable approximation
of the rotation curves of dark matter dominated galaxies. Moreover,
the predicted scaling relation \cite{FM04} fits almost ideally
astronomical observations. In Ref. \cite{SFM06} we predicted the
effects of such scalar field halos for microlensing.

\subsection{Free field}
In the string landscape\cite{LU06}, a triple unification of
inflation, DM and DE suggests itself based on a simple quadratic
potential (\ref{Urot}) with $\chi=0$ corresponding to a free massive
field. Then, our scalar field toy model likewise incorporates `dark
energy' in a rather novel way: The {\em exact parametric solution}
of the equivalent nonlinear Lagrangian $L(R)$ for the `free' field
reads
\ba 
R &=& 6 m^2 x e^x (1 + x )
 , \label{comp21} \\
L &=& \frac{3 m^2}{2\kappa} x e^{2x}( 2 + x)
, \label{comp23}
\ea
where $x:=\ln\Omega$ under the reality condition $\Omega>0$.

\begin{figure}[ht]
\centering \leavevmode\epsfysize=5cm \epsfbox{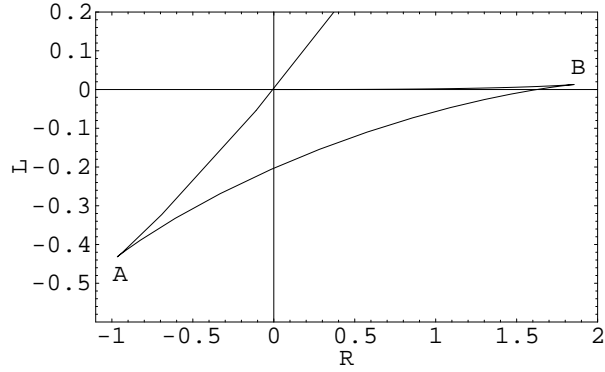}\\% \vskip 0.5cm
\caption{\label{fig2} The swallow tail behavior of the
Lagrangian $L(R)$ for a massive `free' field ($\chi=0$ and
$\kappa=m=1$).}
\end{figure}

The dependence of $L(R)$ given in Fig.~\ref{fig2} is rather
surprising and represents the bifurcation set of the {\em swallow
tail} catastrophe \cite{Ar02,Ku89,kusm} associated with some higher
dimensional grand manifolds. The scalar field $\phi$, its mass $m$
as well as $\chi$ play the role of {\em control parameters}.
According to the theory of singularities (more widely known as {\em
catastrophe theory}, cf.~Arnol'd \cite{Ar02}), this bifurcation set
indicates that the Lagrangian manifolds  are associated with two
local ``minima" and one ``maximum" (and saddle points at the meeting
points of the grand manifold). Each of the ``minima" merges with the
``maximum" at the cuspoidal points A and B and then disappears. The
minima (the semi-infinite segments A and B) are characterized by a
positive second derivative of the Hamiltonian ${\mathcal H}(\Omega)$
with respect to the momentum $\Omega$, i.e.~$d^2 {\mathcal H}/
d\Omega^2>0$. They correspond to an effective Lagrangian with
vanishing cosmological constant.

For the ``maximum", i.e.~the segment AB, this
derivative is negative and the effective Lagrangian has a modified
gravitational constant and a positive cosmological constant,
\be
\kappa_{\rm eff}=e \kappa\,,\qquad \Lambda_{\rm eff}= 3m^2/(2
e)\,,
\ee
respectively.

\begin{figure}[ht]
\centering \leavevmode\epsfysize=7cm \epsfbox{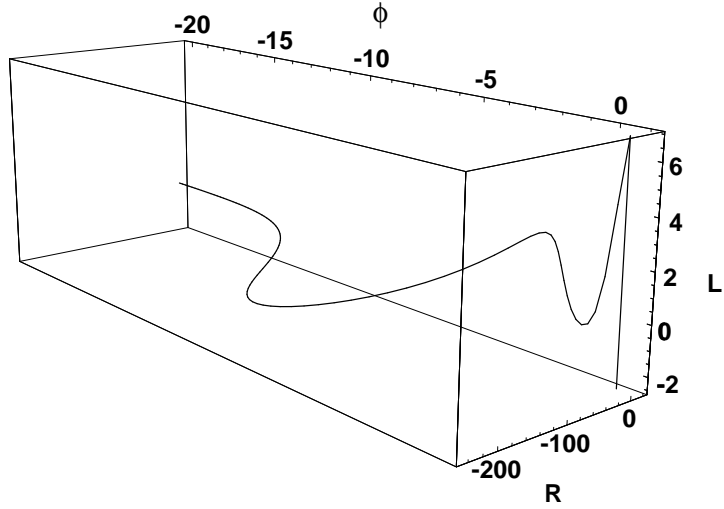}\\ %\vskip 0.5cm
\caption{\label{fig3} Whitney surface for  the nonlinear
curvature Lagrangian $L(R)$ corresponding to the model (\ref{Urot}) with
$\chi=0.3$ and $m=\kappa=1$
depending on $\phi$ as control parameter. ($\phi$ in units of $\left [\sqrt{3/(2\kappa)}\right]$.)}
\end{figure}

\subsection{Self-interacting scalar}
\begin{figure}[ht]
\centering \leavevmode\epsfysize=5cm \epsfbox{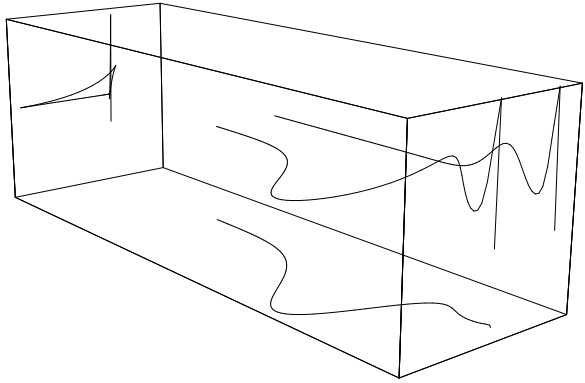}\\ %\vskip 0.5cm
\caption{\label{fig3a}
Projections of the Whitney surface in Fig.~\ref{fig3} for the nonlinear
curvature Lagrangian $L(R)$  corresponding to a nonlinear scalar (\ref{Urot})
with $\chi=0.3$ and $m=\kappa=1$.}
\end{figure}

In the nonlinear case, the {\em exact parametric solution} $L(R)$ reads
\ba
R &=& 6 m^2 x e^x\left [ 1 + x
 - \frac{27\chi}{4\kappa^2} x^4 - \frac{9\chi}{4\kappa^2} x^5 \right ]
 , \label{comp211} \\
L &=& \frac{3 m^2}{2\kappa} x e^{2x}\left [ 2 + x
 - \frac{27\chi}{2\kappa^2} x^4 - \frac{9\chi}{4\kappa^2} x^5 \right ]
.\label{comp231} 
\ea 
The resulting extremal curve of the Whitney
surface is drawn in Fig.~\ref{fig3}, and  its projections $L=L(R)$,
$R=R(\phi)$, and $L=L(\phi)$  in Fig.~\ref{fig3a}.

\begin{figure}[ht]
\centering \leavevmode\epsfysize=6cm \epsfbox{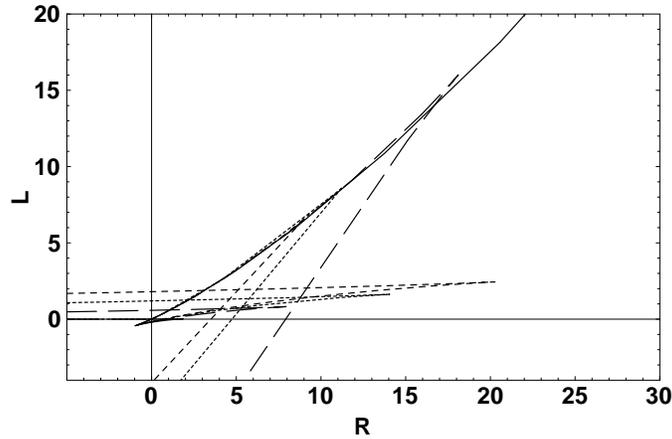}\\% \vskip 0.5cm
\caption{\label{fig4} The wigwam singularity for the nonlinear curvature
Lagrangian $L(R)$ for a NTS model with $\chi=0$, 0.1 (long dashes),
0.2 (short dashes), 0.3 (dashed) and $\kappa=m=1$;
here only the butterfly part near the center is shown.}
\end{figure}

\begin{figure}[ht]
\centering \leavevmode\epsfysize=5cm \epsfbox{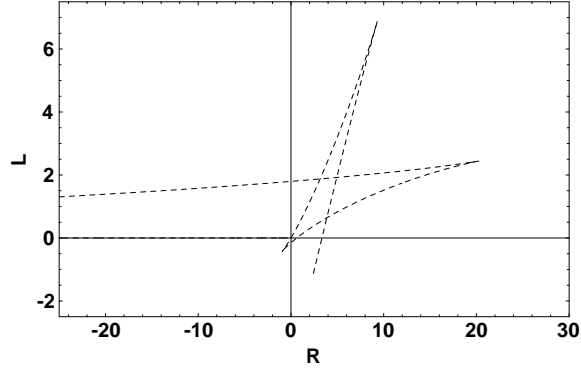}\\% \vskip 0.5cm
\caption{\label{fig5} For the NTS model with $\chi=0.3$ and
$\kappa=m=1$, the  {\em wigwam catastrophe} of the Lagrangian $L(R)$
can be decomposed into a butterfly  plus additional elementary
catastrophe. Here, the butterfly catastrophe arises near the origin;
at about $(R,L)=(-250,-0.015)$ there is a further cusp, cf.~Fig.~\ref{fig6}; the curve
finally returns to the origin which cannot be seen due to our
scale.}
\end{figure}

\begin{figure}[ht]
\centering \leavevmode\epsfysize=5cm \epsfbox{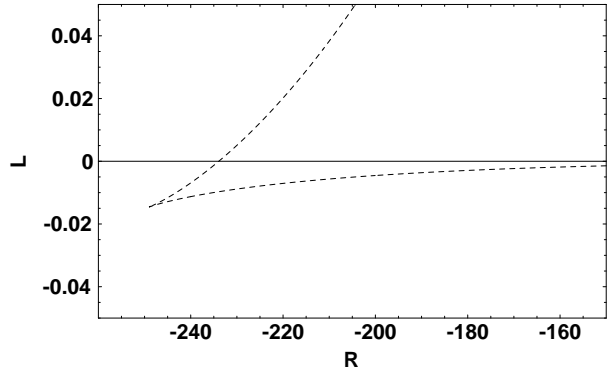}\\% \vskip 0.5cm
\caption{\label{fig6} Continuation of Fig. \ref{fig5}.
Enlarging the additional cusp for the nontopological
soliton field with $\chi=0.3$ and $\kappa=m=1$.}
\end{figure}

{}For $\chi\neq 0$, the bifurcation diagram in Fig.~\ref{fig4}
is of higher rank than for the free field. Near the center, we find the butterfly part of the
catastrophe. But, for each non-vanishing $\chi$, there exists a
further cusp far away from the center, for a large negative $R$
value and $L$ close to zero. From this cusp, the curve finally returns to the origin,
for $L$ values very close to zero so that it cannot be seen in Fig.~\ref{fig4}
or \ref{fig5}.
In an enlargement of the additional cusp in Fig.~\ref{fig6}, one can see that $L(R)$ actually
becomes negative and forms a cusp.

The effective strength of gravity and the value of cosmological
constant 
\be 
\kappa_{\rm eff}=\kappa e^{-x}\, , \quad \Lambda_{\rm
eff}= 3m^2\frac{ x^2 e^x}{x+3}\, ,\label{effcos} 
\ee 
respectively, depend now on
both,  the mass $m$ and, via $x$ as a solution of $R=0$, on the
coupling constant $\chi$ of the scalar field, and need to be
constrained by cosmological data.

In quintessence models \cite{We01}, e.g., the crossover scale of the
scalar field is $\vert\phi_{\rm c}\vert=\exp(-1/\alpha)M_{\rm
Pl}^2/m$, where $\alpha=1/138$ is about Sommerfeld's fine structure
constant and $M_{\rm Pl}=1/\sqrt{\kappa}$ the reduced Planck mass.
Then the tiny observed cosmological constant of the present epoch
\be 
\Lambda_{\rm DE}= m^2\phi_{\rm c}^2=\exp(-2/\alpha)M_{\rm Pl}^4
\simeq (10^{-3}{\rm eV})^4 
\ee 
is roughly reproduced for small $\chi$.

It is important to stress that within the range limited by cuspoidal
points, all three states may coexist with each other. Thus
our bifurcation set, cf. Ref. \cite{SKM05} for a generalization with
$\chi \neq 0$, indicates that each {\em local patch} of the Universe
may have a different strength and cosmological constant (`dark
energy') controlled by the mass  $m$ of the scalar field, but the
effective Lagrangian has approximately  the same Einsteinian form
(\ref{Eineff}). In ``maximal" domains, inflation may be still going
on, thus realizing prospective ideas of Linde \cite{linde}.

\subsection{Natural inflation from the axion?}
In QCD, after integrating out the fermion fields,
its generating functional  including a topological Pontrjagin term for the
Yang-Mills gauge fields induces an effective axion  potential
\be
U = \Lambda_{\rm QCD}^4  \left [1 -\cos (\phi /f_\phi)\right ].
\label{Uaxion}
\ee

This potential displays a periodicity  with a period of $2\pi
f_\phi$, has a minimum at $\phi=0$, as required, and leads to the
induced axion mass of $m_\phi=\Lambda_{\rm QCD}^2/f_\phi$. A
quintaxion may also be induced by spacetime torsion\cite{MR06}.

Within natural inflation\cite{SFK06}, such a  potential has been
proposed for an axion  coupling constant close to the Planck scale.
For simplicity let us assume in the following that
$f_\phi\simeq\sqrt{3/(2\kappa)}$. Then, following the general
prescription (\ref{9.4})-(\ref{9.5}), we find the reparametrized
solution 
\ba 
R &=& 2\kappa \exp \left ( x \right ) \Lambda_{\rm
QCD}^4
 \left [ 2 - 2 \cos \left ( x \right )
   +  \sin \left ( x \right ) \right ]
 \; , \label{nat1} \\
L &=& \exp \left ( 2x \right ) \Lambda_{\rm QCD}^4
 \left [ 1 -  \cos \left ( x \right )
   + \sin \left (x \right ) \right ]
 \; . \label{nat2}
\ea

\begin{figure}[ht]
\centering \leavevmode\epsfysize=5cm \epsfbox{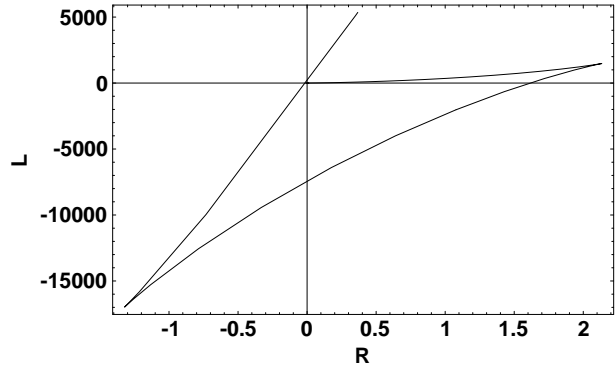}\\% \vskip 0.5cm
\caption{\label{fig8} The axion gives rise again to a swallow tail
catastrophe in the Lagrangian $L(R)$, with the choice of $\Lambda_{\rm
QCD}=0.1$, $\kappa=1$, and $f_\phi=1$.}
\end{figure}

Fig.~\ref{fig8} exhibits a  swallow tail behavior similar as for a NTS
potential with $\chi=0$.

%%%%%%%%%%%%%%%%%%%%%%%%%%%%%%%%%%%%%%%%%%%%%%%%%%%%%%%%%%%%%%
\section{Domain structure of dark matter and
induced dark energy}

Baryonic particles (such as protons and neutrons) account for
visible matter in the Universe, that is only a small fraction of the
observed total matter. The major part of it is a mysterious DM,
which only interacts gravitationally. On the other hand according to
our model, the Universe  splits into almost Einsteinian domains with
-- depending on the physical scale and eras --  a different
gravitational strength and cosmological constant, the latter being
most likely a representation of DE. A natural question is how these
domains arise? To answer this, we have to look into the initial
inflationary stage of the Universe, long before decoupling of
radiation and matter, even before grand unified phase transitions
where only an initial pre-field  existed. At the initial stage of
the quantum era, the fundamental pre-field  may have existed in a
form of a scalar or vector gauge fields. Without lack of generality
we can limit ourselves to a scalar field $\phi$.

During this quantum era, the scalar field was strongly fluctuating.
Due to  the rapid inflation, the size of the Universe as well as the
size of each individual fluctuation have increased enormously while
the amplitude remains almost the same.  During this fast spacelike
expansion, just after the stretching, these space fluctuations of
$\phi$ were frozen while loosing their quantum nature due to the
increasing size.  The decoupling of radiation and  visible matter
has imprinted fluctuations on the electromagnetic spectrum which we
observe now in the anisotropy microwave background radiation.
However, we cannot see such ripples of the visible matter after the
decoupling. Due to the gravitational interaction  the process of
large scale formation has started just after the decoupling. And the
state of nearly homogeneously distributed visible matter existing
after the decoupling has eventually been transformed into clusters
of galaxies and star structures which we see nowadays. Obviously,
the decoupling of the dark part of matter from radiation and visible
matter may have happen even earlier, when the Universe was in the
form of a quark-gluon plasma. However, after decoupling the dark
part of matter must be subjected to the same gravitational forces
and to a similar large scale formation. Since its total mass is
significantly larger than the visible mass, the formation of large
structures of DM may have a weak influence on the visible matter.

Dark matter will have an opposite effect by dictating where and how the
visible matter should  be distributed, probably localized around  clusters of DM.
In the case of scalar field DM, its peculiarities such as  self-interaction
may also have contributed to the DE distribution.
DM cannot be seen directly by traditional observations
but can be inferred from more sophisticated gravitational lensing\cite{Ma07},
revealing that DM exists in a form of a loose network of filaments, growing
over time, which intersect in massive structures at the locations of
clusters of galaxies. This finding is consistent with the conventional
theory of large scale structure formation, where a smooth distribution
of DM collapses first into filaments and then into clusters, forming a
{\em cosmic scaffold}\cite{Ma07}, then  accumulating  visible matter
and  later newly born  stars.

The primary candidate for  DM  is a scalar field. If a scalar field
exists, then  it may have different amplitudes or different mass
densities in different space regions. Such a distribution of $\phi$
 has been created after the inflation of the Universe. It is
associated with different amplitudes of the frozen fluctuations. The
process of  large scale structure formation is rather similar in
each of these regions. Our results indicate that in each of these
different regions the gravitational constant may vary\cite{SKM05},
depending on the amplitude of the scalar field of the given frozen
space fluctuation. In other words, in each of these regions defined
by the original frozen fluctuations there will arise different
Newtonian potentials. This is consistent with recent
studies\cite{Do06} showing that, during expansion, a  vector field
having two different initial amplitudes will give rise to different
Newtonian potentials.  This difference in turn drives
self-consistently enhanced growth of the density perturbations which
enables tiny perturbations to grow into the large structures we see
today.

\subsection{Cosmic domains}
Our studies indicate that these different Newtonian potentials
correspond to different effective gravitational constants in various
regions of the Universe. If this is true, the data of  recent
observations\cite{Ma07} should be reinvestigated. In regions with
larger gravitational constants one may see that the gravitational
lensing will be stronger and vice versa. We regard areas with
different gravitational potentials as {\em cosmic domains}
characterized by their own gravitational and cosmological constants.
There occurs a coexistence of these domains (which primordially may
be of topological origin\cite{Ki80,Ki02}). On the boundaries of the
domains, changes in the gravitational and cosmological constants are
expected to be abrupt. In spite of the random or fluctuational
origin of these domains, the values of these gravitational  and
cosmological constants take only a few universal numbers,
corresponding to only a finite number of types of domains.

{}For small curvature, each of these states  can be described
approximately by the same effective Lagrangian (\ref{Eineff}) with
different gravitational constant $\kappa_{\rm eff}$ and cosmological
constant $\Lambda_{\rm eff}$ and emerge as a fixed point of the
conformal transformation from one side and universal classes of the
smooth differential mappings from the other side\cite{Ku89}. These
spaces are approximately {\em Einstein spaces}, but of different
gravitational strength as well as with different cosmological
constants. The distribution of such cosmological constants in the
Universe corresponds to  DE induced by the scalar field
distribution. On a very large scale, $\phi$  is homogeneously
distributed and therefore we expect that DE will be distributed
homogeneously, depending on the details of  inflationary cosmology.

\subsection{Domains arising from free fields}
If  DM and  DE were associated with a massive free scalar field gravitationally
coupled or/and associated with axions (see, Sec.  5.3) there will arise three
types of domains:

 1) The first type of domains, I, is described by  Einstein`s theory with
conventional gravitational constant and vanishing cosmological constant.

 2) The second type of domains, II, is described by an Einsteinian model
having a very large gravitational constant and a vanishing cosmological constant.
Such type of domains may be gravitationally unstable, leading eventually
to their contraction and ultimately to a gravitational collapse and phenomena
resembling supernova explosions. For very large masses, we may expect that these
explosion will be stronger than  standard supernovae. Our approach may explain
the fact why some supernovae progenitors seem to have exceeded\cite{An06}
the Chandrasekhar limit.

 3) The third type of domains, III, are described by an Einsteinian model with
large gravitational and positive cosmological constant $ \Lambda^{\rm III}_{\rm eff}$
given by (\ref{effcos}). Then we expect that such domains are still expanding
according to  effective Friedman equations and probably their sizes are also increasing.
Such expansion may be ended abruptly and such domains will be transformed into
domains of type I.

One may speculate that  the largest part of the Universe
is occupied by domains of type I,  corresponding to spaces with zero cosmological constant.
The next largest proportion of the Universe is expected to be occupied by
domains of type III. Those parts of the Universe
occupied by domains of type II are decreasing with time.
Each disappearance of such domain would be accompanied by  supernova type explosions.

On the average, the cosmological constant inferred from recent
observations is \be \Lambda_{\rm DE}= \frac{ V_{\rm III}
\Lambda^{\rm III}_{\rm eff} }{ V_{\rm I}+V_{\rm II}+V_{\rm III}},
\ee where the spacelike volumes $V_{\rm I}, V_{\rm II}$ and $V_{\rm
III}$ are associated with  domains of type I, II, and III,
respectively.

In summary, our bifurcation set (see Fig.~\ref{fig1} and
Fig.~\ref{fig2}) indicates that the Universe is locally described by
Einstein's GR and effectively split into domains. The splitting
originated from primordial quantum fluctuations which were frozen
during inflation. Each domain or local patch of the Universe may
have different gravitational strength and may have zero or  negative
pressure associated with DE. In domains of type II associated with a
positive cosmological constant the inflation may be still going on.
It will be stopped exactly at the bifurcation point A and this
domain will be transformed into a domain of type I, as in Fig.
\ref{fig2}. For a  massive free scalar field and for axions the
bifurcation set is the {\em swallow tail} catastrophe. There are two
cusps at  the points A and B which are associated with the highest
singularities of the differential mappings. They are also related to
the domain boundaries. In each of these cuspoidal points A and B of
the bifurcation diagram, the minimum of some grand manifold merges
with the maximum\cite{Kus}. Each local minimum (the segment A and
the semi-infinite segment B) is associated with two types of
domains, I and II, respectively. Each local maximum (saddle) of the
grand manifold is associated with an expanding domain of type III
(the segment AB). The effective Einsteinian gravity for this domain
is stronger than that in the domain of type I, depending on the mass
of the scalar field. Domains with a positive effective cosmological
constant are still in an expanding inflationary phase.

The `strong' gravity state, the domain II (the semi-infinite segment
B) with a negative or vanishing cosmological constant corresponds to
deflation. Since the Universe is split into dynamical regions
associated with different gravity,  their boundaries change with
time and present some kind of cosmological strings or membranes.
Changes of the boundary are associated with some sort of local phase
transition reminding us of phase separations like those arising in
first order phase transitions. At  some instant, all these domains
are  in quasi-equilibrium with each other although their boundaries
changes with time as the membranes are moving. At very long time
scales (much larger than the Planck time) when inflation in the
domains of type III will  eventually be ceased, only stable phases,
like the domains of type I and some small concentration of domains
of type II will remain. These domains would produce explosions
resembling supernovae with  masses well beyond of the Chandrasekhar
limit, similar to those observed\cite{An06}.

%********************************************
\subsection{Domains arising from self-interacting fields}
In case of self-interacting  scalar fields with $\chi\neq 0$, the
classification  will be richer (see Figures \ref{fig3a}, \ref{fig4}
and \ref{fig5}), with {\em five} types of domains. The bifurcation
diagram associated with such a {\em wigwam} catastrophe or $A_5$
singularity according to Arnold's classification\cite{Ar02} consists
of  four cusps or four $A_2$ singularities, with a dramatic change
in the gravitational and cosmological constants.

There occurs a  general decoupling of
the wigwam catastrophe $A_5$ into a butterfly catastrophe $A_4$,
and the elementary cusp $A_2$, depending on the values of $R$:
If the butterfly catastrophe arises near the origin
or at small values of $R$, the additional cuspoidal point is
associated with a very large value of $R\simeq -250$.
For some fixed values of the coupling constant $\chi=0.1$, 0.2, 0.3,
the butterfly part of bifurcation diagram is presented in Figs.
\ref{fig4}. In comparison with the free massive field, the inclusion
of the self-interaction associated with the nonzero parameter $\chi$
gives rise to two extra cusps leading to the appearance of two extra
branches in the bifurcation diagram, again associated with effective
Einsteinian spaces. Although these almost linear branches do exist
at positive $R$, their major part is related to negative values of
$R$. For one of these branches the cosmological constant vanishes while for the
other one  the cosmological constant decreases when the
value of $\chi$ increases, and  the part of the branch associated
with  $R>0$ decreases. In general, the self-interaction induced by the
field $\phi$ increases the strength of gravity  and decreases the
cosmological constant.

Let us describe in more detail the {\em transitions} in the
bifurcations:
Starting from the first branch, denoted as I, through the first cuspoidal
point we arrive at the second branch II which is very similar to the
case of a free massive scalar field, although its length decreases
significantly when $\chi$ increases and the cosmological constant
vanishes at any value of $\chi$. This branch corresponds to the
limit  $\phi \rightarrow 0$ of small scalar fields, corresponding to
the conformal factor $\Omega =1$. Therefore, on this branch  we
recover the linear Hilbert-Einstein Lagrangian $L_{\rm HE}
=R/2\kappa$, as expected. With decreasing $\Omega$, we will come to
the  next cuspoidal point, where the new branch III  has a positive
cosmological constant and  gravity becomes stronger. With increasing
value of $\chi$ the length of this branch increases. Thus, there
will be more domains having positive cosmological constant and a
gravitational strength higher than the conventional. By decreasing
$\Omega$ until reaching  the next cuspoidal point of the bifurcation
diagram, we arrive at the branch IV. The spacetime associated with
this branch has a very strong gravity and negative cosmological
constant. Similar to the free massive case, this branch
probably is gravitationally unstable.
Finally, for very small $\Omega$, we will arrive at the branch
V with $R<0$, where  gravity is the strongest but without
cosmological constant. Similar to the previous branch IV, the domain
V  is unstable, as well. Thus  in a Universe filled with
self-interacting scalar fields, we would expect that a rather strong
phase separation into further domains will arise.

%************************************
\section{Discussion}
For dilute DM the Lagrangian has the standard Hilbert-Einstein form
$L_{\rm HE} =R/2\kappa$. For dense DM, i.e. large values of the DM
scalar, the resulting effective Lagrangian is $L_{\rm HE}
=R/2\kappa_{\rm eff}$, only the slope is less steep, i.e. the
effective gravitational coupling $\kappa_{\rm eff}> \kappa$ is, by
``renormalization" \cite{We01,Be88}, larger than Newton's.  This
resembles some aspects of MOND, but with the advantage that we still
work in the standard general relativistic framework. Depending on
its sign, the cosmological constant  $\Lambda_{\rm eff}$  in a
bifurcation  approximated for small $R$ by (\ref{Eineff}) can also
model DE or an accelerating phase of the present epoch of the
Universe (like `anti-gravity'). There also may arise ``gravitational
screening" for $\kappa_{\rm eff}$ smaller than Newton's. MACHOs can
now be described by some specific nonlinear modifications of the
Hilbert-Einstein action.

In the light of our finding we have to comment on recent
observations\cite{Me07}, which constrain  the density of MACHO type
objects  in the Universe by measuring the brightness of high
redshift type Ia supernovae relative to  low redshift samples. These
data favor DM made of microscopic particles (such as WIMPs) or
scalar  and vector gauge fields  over MACHOs with masses between
$10^{-2}$ and $10^{10}$ solar masses. This  provides another
evidence that our approach correctly describes the frozen space
fluctuations which lead to a formation of cosmic domains with a
different density of DM and DE in the Universe.

\footnotesize

\section*{Acknowledgments}
We would like to thank Burkhard Fuchs, Humberto Peralta, and
Konstantin Zioutas for helpful discussions. One of us (EWM) thanks
Noelia,  Markus G\'erard Erik, and Miryam Sophie Naomi for
encouragement.

%\vfill
%\eject

%%%%%%%%%%%%%%%%%%%%%%%%%%%%%%%%%%%%%%%%%%%%%%%%%%%%%%%%%%%%%%%%%%%%%

\vfill
%\pagebreak
\end{document}